\documentclass[PRD,twocolumn,showpacs,preprintnumbers,nofootinbib,amsmath,amssymb]{revtex4}

\usepackage{graphicx}
\usepackage{dcolumn}
\usepackage{bm}

\def\fnl{{f_{\rm{nl}}}}

\def\hnl{{f_{\rm{nl}}}}
\def\hnle{{\widehat \hnl}}
\def\hnlb{{\widehat{\hnl^b}}}
\def\hnlett{{\widehat{(\hnl^2)^t}}}
\def\epsilone{{\widehat \epsilon}}

\def\VEV#1{\left\langle #1 \right\rangle}

\newcommand{\beq}{\begin{equation}}
\newcommand{\eeq}{\end{equation}}
\newcommand{\beqa}{\begin{eqnarray}}
\newcommand{\eeqa}{\end{eqnarray}}

\newcommand{\Var}{{\mathrm{Var}}}
\begin{document}

\title{The CMB Bispectrum, Trispectrum, non-Gaussianity, and the
Cramer-Rao Bound}

\author{Marc Kamionkowski}
\affiliation{California Institute of Technology, Mail Code 350-17,
Pasadena, CA 91125}
\author{Tristan L.~Smith}
\affiliation{Berkeley Center for Cosmological Physics, Physics
     Department, University of California, Berkeley, CA 94720} 
\author{Alan Heavens}
\affiliation{SUPA, Institute for Astronomy, School of Physics,
     University of Edinburgh, Royal Observatory, Blackford Hill,
     Edinburgh, EH9 3HJ, U.K.} 

\date{\today}

\begin{abstract}
Minimum-variance estimators for the parameter $\fnl$ that
quantifies local-model non-Gaussianity can be constructed from
the cosmic microwave background (CMB) bispectrum (three-point
function) and also from the
trispectrum (four-point function).  Some have suggested that a
comparison between the estimates for the values of $\fnl$ from
the bispectrum and trispectrum allow a consistency test for the
model.  But others argue that the saturation of the
Cramer-Rao bound by the bispectrum estimator implies that no
further information on $\fnl$ can be obtained from the
trispectrum.  Here we elaborate the nature of the
correlation between the bispectrum and trispectrum
estimators for $\fnl$.  We show that the two estimators become
statistically independent in the limit of large number of CMB
pixels and thus that the trispectrum estimator {\it does} indeed
provide additional information on $\fnl$ beyond that obtained
from the bispectrum.  We explain how this conclusion is
consistent with the Cramer-Rao bound.  Our discussion of the
Cramer-Rao bound may be of interest to those doing Fisher-matrix
parameter-estimation forecasts or data analysis in other areas
of physics as well.
\end{abstract}


\maketitle

\section{Introduction}

Observations of the cosmic microwave background (CMB) have
confirmed a now `standard' cosmological model \cite{Komatsu:2010fb}.  A
key aspect of this model is that primordial fluctuations 
are a realization of a Gaussian random field.  This implies that
CMB fluctuations are completely characterized by their two-point
correlation function $C(\theta)$ in real space, or equivalently,
the power spectrum $C_{\ell}$ in harmonic space.  All
higher-order $N$-point correlation functions with even $N$ can
be written in terms of the two-point function, and all $N$-point
correlation functions with odd $N$ are zero.

But while the simplest single-field slow-roll (SFSR) inflationary
models assumed in the standard cosmological model predict
departures from Gaussianity to be undetectably small
\cite{localmodel}, several beyond-SFSR models predict departures
from Gaussianity to be larger \cite{larger},
and possibly detectable with current
or forthcoming CMB experiments.  While the range of predictions
for non-Gaussianity is large, the local model for
non-Gaussianity \cite{Luo:1993xx}---that which
appears in arguably the simplest beyond-SFSR models---has become
the canonical model for most non-Gaussianity searches.  The
non-Gaussianity is parametrized in these models by a
non-Gaussian amplitude $\fnl$ to be defined more precisely below.

Most efforts to measure $\fnl$ have relied on an estimator
constructed from the CMB bispectrum, the three-point correlation
function in harmonic space.  However, the local model also
predicts a non-zero trispectrum (the harmonic-space four-point function)
\cite{Kunz:2001ym,Hu:2001fa,Okamoto:2002ik,Kogo:2006kh,Regan:2010cn},
and efforts have recently been mounted to determine $\fnl$
from the trispectrum \cite{Smidt:2010ra}.   It has been
suggested, moreover, that a comparison of the values of $\fnl$
obtained from the bispectrum and trispectrum can be used as a
consistency test for the local model \cite{Byrnes:2006vq,
Kogo:2006kh,Smidt:2010ra}.

However, it can be shown that the bispectrum estimator for
$\fnl$ saturates the Cramer-Rao bound, and it has been argued
that this implies that no new information on the value of
$\fnl$, beyond that obtained from the bispectrum, can be
obtained from the trispectrum \cite{Babich:2005en,Creminelli:2006gc}.
Ref.~\cite{Creminelli:2006gc} further outlines the nature of
the correlation between the bispectrum and trispectrum $\fnl$
estimators implied by this conclusion.

Here we show that the trispectrum {\it does} provide additional
information on $\fnl$; i.e., it is {\it not} redundant with that
from the bispectrum.  We show that there is indeed a
correlation between the bispectrum and trispectrum $\fnl$
estimators, elaborating the arguments of
Ref.~\cite{Creminelli:2006gc}.  However, we show with analytic
estimates and numerical calculations that this correlation
becomes weak in the high-statistics limit.  We explain, with a
simple example, how additional information on $\fnl$ can be
provided by the trispectrum given that the bispectrum estimator
for $\fnl$ saturates the Cramer-Rao bound.  Put simply, the
Cramer-Rao inequality bounds the variance with which a
distribution can be measured, but there may be additional
information in a distribution, about a theory or its parameters,
beyond the distribution variance.  The discussion of
the Cramer-Rao bound and the examples we work out in
Section~\ref{sec:cramerrao} may be of interest to a much broader
audience of readers than just those interested in CMB
non-Gaussianity.

The outline of this paper is as follows:  We begin in Section
\ref{sec:cramerrao} with our discussion of the Cramer-Rao
bound.  The aim of the rest of the paper is to illustrate
explicitly the nature of the correlation between the bispectrum
estimator for $\fnl$ and the trispectrum estimator for $\fnl^2$
and to show that the correlation becomes small in the high-statistics limit.
In Section \ref{sec:definitions} we introduce our conventions for the
bispectrum and trispectrum.  In Section \ref{sec:estimators} we
derive the minimum-variance estimators for $\fnl$ from the
bispectrum and trispectrum and evaluate the noises in each.  We
also write down approximations for the estimators and noises
valid for the local model.  In Section
\ref{sec:crosscorrelation} we explain the nature of the
correlation between the bispectrum and trispectrum
estimators for $\fnl$.  We then show that this
correlation becomes weak (scaling with $(\ln
N_{\mathrm{pix}})^{-1}$) as the number $N_{\mathrm{pix}}$ of
pixels becomes large.  We conclude in Section
\ref{sec:conclusion}.  Appendix A details the correspondence
between continuum and discrete Fourier conventions for power
spectra, bispectra, and trispectra, and Appendix B provides
describes the numerical evaluation of the correlation.

\section{The Cramer-Rao Bound}
\label{sec:cramerrao}

In the Sections below we will demonstrate that the estimators
for $\fnl$ and $\fnl^2$ becomes statistically independent with
sufficiently good statistics.  However, the bispectrum estimator
for $\fnl$ saturates the Cramer-Rao bound, and it has been
argued that this saturation implies that no further information
about $\fnl$, beyond that obtained from the bispectrum, can be
obtained from the trispectrum \cite{Babich:2005en,Creminelli:2006gc}.
Here we explain that the Cramer-Rao inequality bounds only the
variance with which $\fnl$ can be measured; additional
information, beyond the variance, can be obtained from
measurement of $\fnl^2$ from the trispectrum.

To illustrate, consider, following
Ref.~\cite{Babich:2005en}, the analogous problem of determining $\fnl$
and $\fnl^2$ from a one-dimensional version of the local model.
Suppose we have a random variable $X$ written in terms of a Gaussian
random variable $x$ of zero mean ($\VEV{x}=0$) and unit variance
($\VEV{x^2}=1$) as
$X=x+\epsilon(x^2-1)$.  Here, $\epsilon$ parametrizes the
departure from the null hypothesis $\epsilon=0$.  The PDF for
$X$, for a given $\epsilon$, is
\begin{equation}
    P(X|\epsilon) = \frac{1}{\sqrt{2\pi}} \left[
    \frac{e^{-x_+^2/2}}{1+2 \epsilon x_+}+ \frac{e^{-x_-^2/2}}{1+2
    \epsilon x_-} \right],
\label{eqn:onedPDF}
\end{equation}
where
\begin{equation}
     x_{\pm} = \frac{1}{2\epsilon} \left[ \pm \sqrt{1+4
     \epsilon(X+\epsilon)}-1 \right].
\label{eqn:xpm}
\end{equation}
The logarithm of the PDF can then be Taylor expanded about
$\epsilon=0$ as
\begin{equation}
     \ln P(X|\epsilon) = -\frac{X^2}{2} + \epsilon I_1(X) -
     \frac{\epsilon^2}{2} I_2(X) + {\cal O}(\epsilon^3),
\label{eqn:expansion}
\end{equation}
where $I_1(X) \equiv X^3-3X$, and $I_2(X) = 5X^4+5-14 X^2$.  It
will be useful below to note that the expectation values of
these quantities in the weakly non-Gaussian limit are
$\VEV{I_1}=6\epsilon + {\cal O}(\epsilon^3)$ and $\VEV{I_2} = 6
+ 272\,\epsilon^2 +{\cal O}(\epsilon^4)$.

Now suppose we have a realization consisting of $N$ data
points $X_i$, each drawn independently from the distribution in
Eq.~(\ref{eqn:onedPDF}), and let's arrange these data points into
a vector $\mathbf{X}$.  The PDF for this realization, for a
given $\epsilon$, is
\begin{eqnarray}
     \ln P(\mathbf{X}|\epsilon) &=& \sum_i\left[ - \frac{X_i^2}{2} + \epsilon
     I_1(X_i) \right. \nonumber \\
     & & \left. - \frac{\epsilon^2}{2} I_2(X_i) + {\cal
     O}(\epsilon^3) \right].
\label{eqn:multiP}
\end{eqnarray}

The Cramer-Rao inequality states that the smallest variance
$\mathrm{Var(\epsilone)} \equiv \VEV{\epsilone^2}-\VEV{\epsilone}^2$ to an
estimator $\epsilone$ is
\begin{equation}
     \mathrm{Var(\epsilone)} \geq \frac{1}{F},
\label{eqn:CRbound}
\end{equation}
where
\begin{eqnarray}
     F &=& \int \left[ \frac{\partial \ln
     P(\mathbf{X}|\epsilon)}{\partial \epsilon}
     \right]^2 \,P(\mathbf{X}|\epsilon) d\mathbf{X} \nonumber \\
     &\equiv& \VEV{
     \left[ \frac{\partial \ln P(\mathbf{X}|\epsilon)}{\partial
     \epsilon} \right]^2},
\label{eqn:Fisher}
\end{eqnarray}
is the Fisher information.  Here, the angle brackets denote an
expectation value with respect to the null-hypothesis
($\epsilon=0$) PDF.  Applying Eq.~(\ref{eqn:Fisher}) to
Eq.~(\ref{eqn:multiP}), we find
\begin{equation}
     F = \sum_i \VEV{[I_1(X_i)]^2} = 6N,
\label{eqn:Fone}
\end{equation}
from which we infer
\begin{equation}
     \mathrm{Var(\epsilone)} \geq \frac{1}{6\,N}.
\label{eqn:CRforf}
\end{equation}

This model predicts a skewness $\VEV{I_1} = \VEV{X^3-3X} =
6\epsilon$, and so we can construct an estimator for $\epsilon$ from the
measured skewness as follows:
\begin{equation}
     \epsilone_{s} = \frac{1}{6N} \sum_i (X_i^3-3 X_i).
\label{eqn:skewestimator}
\end{equation}
The variance to this estimator is $\mathrm{Var(\epsilone_{s})} =
(6N)^{-1}$, and so this estimator saturates the Cramer-Rao bound.

In retrospect, this saturation should come as no surprise.
According to Eqs.~(\ref{eqn:multiP}) and (\ref{eqn:Fisher}), the
Fisher information---and thus the minimum variance with which
$\epsilon$ can be measured---is determined entirely by the term in
$\ln P(\mathbf{X}|\epsilon)$ linear in $\epsilon$ which, in this case,
is precisely the skewness.  Thus, the
terms in $\ln P(\mathbf{X}|\epsilon)$ that are higher order in
$\epsilon$ contribute nothing to the Fisher information.  And
since the term linear in $\epsilon$ multiplies the skewness,
$\epsilone_s$ saturates the Cramer-Rao bound.

\begin{figure}[htbp]
\centering
\includegraphics[width=0.48\textwidth]{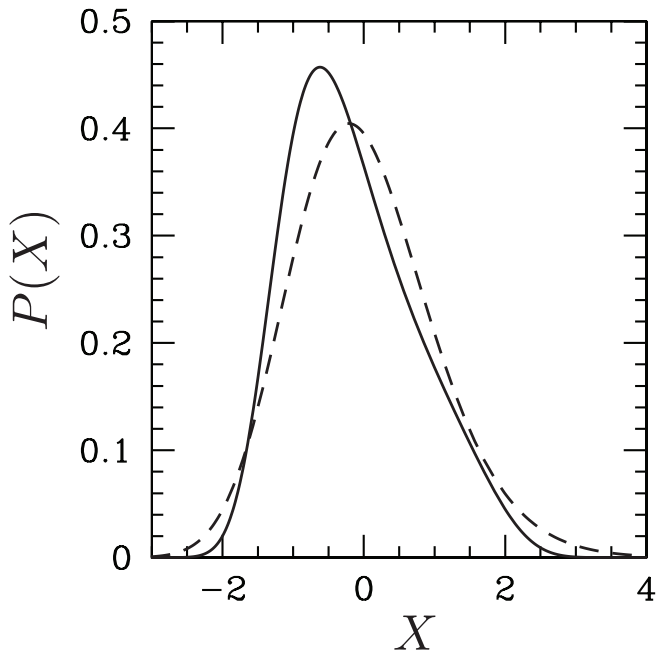}
\caption{Here we plot two probability distribution functions
     that share the same skewness but with two different values
     for the kurtosis.}
\label{fig:samplePDFs}
\end{figure}

But this does {\it not} mean that there is no information about
$\epsilon$ from these higher-order terms.  Consider, for
example, a more general PDF,
\begin{eqnarray}
     \ln P_\alpha(X|\epsilon,\epsilon_1^2) &=& -\frac{X^2}{2} + \epsilon
     I_1(X) \nonumber \\
     & & -  \frac{\epsilon_1^2}{2} I_2(X) + {\cal O}(\epsilon^3),
\label{eqn:alphaexpansion}
\end{eqnarray}
parametrized by $\epsilon_1^2$, in addition to the parameter
$\epsilon$.  This PDF differs from the PDF in
Eq.~(\ref{eqn:expansion}) in the coefficient of $I_2(X)$.
In the weakly non-Gaussian limit, the skewness of this PDF is
$\VEV{I_1(X)} = 6\epsilon$, and its ``kurtosis'' is
$\VEV{I_2(X)} = 6+846\,\epsilon^2 -574\,\epsilon_1^2+
18(\epsilon^2-\epsilon_1^2)$.\footnote{In this paper we use the
term ``kurtosis'' to denote the expectation value of $I_2(X)$.
This is qualitatively similar to, but slightly different, than
the usual kurtosis, which is usually defined to be the
expectation value of $X^4-6X^2+3$.}  If we fix $\epsilon$, we
then have a family of PDFs, parametrized by $\epsilon_1$, that
all have the same skewness but with different values of the
kurtosis.  Fig.~\ref{fig:samplePDFs} shows two PDFs that have
the same skewness but different kurtoses.  These are clearly two
very different distributions; qualitatively, the large-$X$ tails
are suppressed as $\epsilon_1$ is increased.

The estimator in Eq.~(\ref{eqn:skewestimator}) once again gives
us the optimal estimator for $\epsilon$ in this new PDF, but we
can now also measure from the data the kurtosis, the expectation
value of $I_2(X)$, which provides an estimator for
$846\,\epsilon^2 -574\,\epsilon_1^2+
18(\epsilon^2-\epsilon_1^2)$.  This
can then be used in combination with the skewness
estimator for $\epsilon$ to obtain an estimator for $\epsilon_1^2$.
According to the Cramer-Rao inequality, the smallest variance to
$\epsilon_1^2$ that can be obtained is
\begin{eqnarray}
     \Var(\epsilon_1^2) &=& \Biggl\{ \int \, \left[\frac{\partial \ln
     P(\mathbf{X}|\epsilon,\epsilon_1^2)}{\partial(\epsilon_1^2)} \right]^2
     \nonumber \\
     & & \times P(\mathbf{X}|\epsilon,\epsilon_1^2)\, dX
     \Biggr\}^{-1} = \frac{1}{278\,N}.
\label{eqn:fnl2}
\end{eqnarray}
Note that we cannot apply the Cramer-Rao bound to the parameter
$\epsilon_1$, rather than $\epsilon_1^2$, as $\partial
P(\mathbf{X}|\epsilon,\epsilon_1^2)/\partial \epsilon_1$ is zero under
the null hypothesis $\epsilon_1=0$, thus violating one of the
conditions for the Cramer-Rao inequality to apply.  Since
$\epsilon_1^2$, not $\epsilon_1$, is determined by the data, the
distribution function for $\epsilon_1^2$ (not $\epsilon_1$) will
approach a Gaussian distribution in the large-$N$ limit.  

The covariance between $\epsilon$ and $\epsilon_1^2$ is zero, as the
former is odd in $X$ and the latter even.  Still, this does not
necessarily imply that the two are statistically independent, as
there is still a covariance between $\epsilon^2$ and
$\epsilon_1^2$.  However, this becomes small as $N$ becomes
large.  The correlation coefficient in this example is $r \equiv
\mathrm{Cov}(\epsilon^2,\epsilon_1^2)/\sqrt{ \Var(\epsilon^2)
\Var(\epsilon_1^2)} \simeq 6\,N^{-1/2}$.  Thus, for large $N$,
$\epsilon$ and $\epsilon_1^2$ are two statistically independent
quantities that can be obtained from the data and then compared
with the local-model prediction that $\epsilon_1^2=\epsilon^2$.
In brief, the skewness and kurtosis are two different quantities
that can be obtained from a measured distribution.  In the
limit of large $N$, no measurement of the skewness, no matter
how precise, can tell us anything about the kurtosis, and {\it
vice versa}.

In this example, a one-sigma excursion
in $\epsilon$ from a measurement with $N$ data points is
$\Var^{1/2}(\epsilon) = (6N)^{-1/2}$, and this is smaller
than $\Var^{1/4}(\epsilon_1^2) = (278\,N)^{-1/4}$, the
square root of the one-sigma excursion in $\epsilon_1^2$, for any
$N\gtrsim$~few.  Thus, the skewness will
provide better sensitivity if we are simply trying to detect
a departure from the null hypothesis $\epsilon=0$; measurement of
$\epsilon_1^2$ will not add much in this case.  Still, if $\epsilon$ is
measured with high statistical significance from the skewness,
then measurement of $\epsilon_1^2$ can, with sufficient statistics,
provide a statistically independent determination of $\epsilon^2$
and/or an independent test of the theory.

Now consider another PDF,
\begin{eqnarray}
     \ln P_{\mathrm{small}}(X|\epsilon) &=& -\frac{X^2}{2} + 10^{-2} \epsilon
     I_1(X) \nonumber \\
     & &  -  \frac{\epsilon^2}{2} I_2(X) + {\cal O}(\epsilon^3),
\label{eqn:epsexpansion}
\end{eqnarray}
that differs from the local-model PDF in the suppression we have
inserted for the term linear in $\epsilon$, which thus suppresses
the skewness.  Application of the
Cramer-Rao inequality in this case tells us that the smallest
value of $\epsilon$ that can be distinguished from the null
hypothesis ($\epsilon=0$) is $10^2/\sqrt{6N}$, and we know from
the discussion above that this variance is obtained via
measurement of the skewness.  However, $\epsilon^2$, the
coefficient of the second term in the expansion---that obtained
from measurement of the kurtosis---can be obtained with the
variance given above.  Thus,
in this case, estimation of $\epsilon^2$ via measurement of the
kurtosis, provides a more sensitive probe of a departure from the null
hypothesis $\epsilon=0$  than does estimation of $\epsilon$ from
measurement of the skewness, as long
as $N\lesssim 10^7$.  Note that the Cramer-Rao bound is not
violated in this case, as measurement of $\epsilon^2$, which does
not discriminate between positive and negative values of $\epsilon$,
does not provide any further information on $\Var(\epsilon)$.  The
{\it apparent} violation of the Cramer-Rao bound arises in this
case because one of the conditions for the validity of the
Cramer-Rao bound---that $\partial \ln P/\partial \epsilon$ be
non-zero at $\epsilon=0$ (under the null hypothesis)---is becoming
invalid as the numerical coefficient of $\epsilon$ in $\ln P$ is
made smaller.  Had we chosen that coefficient to be zero, rather
than $10^{-2}$, then the Cramer-Rao inequality would have given a
nonsensical bound for $\Var(\epsilon)$.

\subsection{Summary}

Suppose we have a theory that
predicts new effects parametrized by a quantity $\epsilon$, with
$\epsilon=0$ representing the null hypothesis.  A general PDF for
the data $\mathbf{X}$ given $\epsilon$ (or likelihood for $\epsilon$ for
given data $\mathbf{X}$) can be expanded in
$\epsilon$ as $\ln P(X|\epsilon) = \ln P_0(X) + \epsilon g(X) + \epsilon^2
h(X)+\cdots$, where $P_0(X)$ is the PDF under the null
hypothesis $\epsilon=0$ and $g(X)$ and $h(X)$ are functions that
describe the theory.  Estimation of $\epsilon$ can be obtained
through measurement of the mean value of $g(X)$, and an
independent estimation of $\epsilon^2$ can, with sufficiently
good statistics, be obtained from
measurement of the mean value of $h(X)$.  If $\VEV{[g(X)]^2}^2 \gtrsim
\VEV{[h(X)]^2}$, where the expectation value is with respect to
$P_0$, then measurement of the mean value of $g(X)$ will provide
a more sensitive avenue for detection of a value of $\epsilon$ that
departs from the null hypothesis than measurement of the mean
value of $h(X)$.  If $\VEV{[g(X)]^2}^2 \lesssim \VEV{[h(X)]^2}$,
then measurement of the mean value of $h(X)$ will provide a more
sensitive test for detection of a value of $\epsilon$ that departs
from the null hypothesis.  If the two are comparable, then both
tests will be comparable.  In the case of a
statistically-significant detection, there may be, given
sufficient statistics, independent information on the values of
$\epsilon$ and $\epsilon^2$ from measurement of both moments.
Care must be taken in interpreting results of measurement of
$\epsilon^2$ from $h(X)$, to note that the distribution of the
$h(X)$ estimator for $\epsilon^2$ is Gaussian in $\epsilon^2$,
not $\epsilon$.

\subsection{Local-model bispectrum and trispectrum}

Similar arguments apply, {\it mutatis mutandis}, to measurement
of the bispectrum and trispectrum, generalizations of the
skewness and kurtosis: the estimator for $\fnl$ obtained from the
bispectrum is statistically independent (for sufficiently
large $N_{\mathrm{pix}}$) from the estimator for $\fnl^2$
obtained from the trispectrum.  If the variance to $\fnl$
obtained from the bispectrum is comparable to the square root of
the variance to $\fnl^2$ obtained from the trispectrum
\cite{Hu:2001fa,Kogo:2006kh}, both
will have roughly comparable sensitivities toward detection of a
departure from the null hypothesis $\fnl=0$.  If there is a
statistically significant detection, both can provide, with
sufficiently good statistics, independent information on $\fnl$
and $\fnl^2$, even if the bispectrum estimator for $\fnl$
saturates the Cramer-Rao bound.  We stop short of verifying
these claims with the full likelihood for the local model.
However, the arguments given explicitly for the one-dimensional
analog above also apply to the skewness and kurtosis in the
local model, the three- and four-point functions at zero lag,
respectively.  While the skewness and kurtosis are not optimal
estimators for $\fnl$ or $\fnl^2$, they are statistically
independent quantities that are derived from the bispectrum and
trispectrum, respectively.

\subsection{Another example}

Here we provide another example where statistically-independent
information can be provided for estimators for $\epsilon$ and
$\epsilon^2$, where $\epsilon$ is a parameter  that quantifies a
departure from a null hypothesis.
Suppose we want to test a theory in which the decay product
from a polarized particle is predicted to have
an angular distribution $P(\theta) \propto P_0(\theta) +
\epsilon P_1(\theta) +\epsilon^2 P_2(\theta)$, where $P_n$ are
Legendre polynomials, and $\epsilon$ parametrizes the departure
from the null hypothesis.  In this case, measurement of the
dipole, the mean value of $P_1(x)$, provides an estimator
for $\epsilon$, and measurement of the quadrupole, the
mean value of $P_2(x)$, provides a statistically-independent
(with sufficiently high statistics) estimator for $\epsilon^2$.
Thus, measurement of both the dipole and quadrupole can be used
to test the data, even though the Cramer-Rao inequality tells us
that $\Var(\epsilon)$ is bounded by the value obtained from the
dipole.

\section{Definitions and Conventions}
\label{sec:definitions}

We have argued above that the bispectrum estimator for $\fnl$
and the trispectrum estimator for $\fnl^2$ may provide
statistically independent information.  The aim of the rest of
the paper will be to evaluate explicitly the correlation
between the bispectrum estimator for $\fnl$ and the trispectrum
estimator for $\fnl^2$.  We will find that it is nonzero, but
that it becomes small in the large-$l_{\mathrm{max}}$ limit.

We assume a flat sky to avoid the complications (e.g., spherical
harmonics, Clebsch-Gordan coefficients, Wigner 3$j$ and 6$j$
symbols, etc.) associated with a spherical sky, and we further
assume the Sachs-Wolfe limit.  We denote the fractional
temperature perturbation at position $\vec\theta$ on a flat sky
by $T(\vec\theta)$, and refer to it hereafter simply as the
temperature.

The temperature in the local model is written,
\begin{equation}
     T(\vec \theta) = t(\vec\theta) +\fnl [t(\vec\theta)]^2,
\end{equation}
in terms of a Gaussian random field $t(\vec \theta)$.  Note that
our $\fnl$ is three times the definition, in terms of the
gravitational potential, used in most of the literature.  We use
this alternative definition to simplify the equations, but the
difference should be noted if comparing our quantitative results
with others.  The field $t(\vec\theta)$ has a power spectrum
$C_l$ given by
\begin{equation}
     \VEV{t_{\vec l_1} t_{\vec l_2}} = \Omega \delta_{\vec
     l_1+\vec l_2,0} C_l,
\label{eqn:powerspectrum}
\end{equation}
where $\Omega=4\pi f_{\mathrm{sky}}$ is the survey area (in
steradian), $t_{\vec l}$ is the Fourier transform of
$t(\vec\theta)$, and $\delta_{\vec l_1+\vec l_2,0}$ is a Kronecker
delta that sets $\vec l_1 = -\vec l_2$.  In the limit $\hnl T
\ll 1$ (current constraints are $\hnl T \lesssim 10^{-3}$), $C_l$
is also the power spectrum for $T(\vec\theta)$.

The bispectrum $B(l_1,l_2,l_3)$ is  defined by 
\begin{equation}
     \VEV{T_{\vec l_1} T_{\vec l_2} T_{\vec l_3}} = \Omega
     \delta_{\vec l_1 +\vec l_2 +\vec l_3,0} B(l_1,l_2,l_3).
\label{eqn:bispectrum}
\end{equation}
The Kronecker delta insures that the bispectrum is defined only
for $\vec l_1 +\vec l_2+\vec l_3=0$; i.e., only for triangles in
Fourier space.  Statistical isotropy then dictates that the
bispectrum depends only on the magnitudes $l_1$, $l_2$, $l_3$ of
the three sides of this Fourier triangle.  The bispectrum for
the local model is,
\begin{equation}
     B(l_1,l_2,l_3) = 2 \hnl [ C_{l_1} C_{l_2} + C_{l_1} C_{l_3} +
     C_{l_2} C_{l_3}].
\label{eqn:lmbispectrum}
\end{equation}

Likewise, the trispectrum is defined by
\begin{equation}
     \VEV{T_{\vec l_1} T_{\vec l_2} T_{\vec l_3}
      T_{\vec l_4}} = \Omega
     \delta_{\vec l_1 +\vec l_2 +\vec l_3+ \vec l_4,0} 
     {\cal T}(\vec l_1,\vec l_2,\vec l_3,\vec l_4),
\end{equation}
and for the local model,
\begin{eqnarray}
          {\cal T}(\vec l_1,\vec l_2,\vec l_3,\vec l_4) &=& \hnl^2 \left[
          P_{l_3l_4}^{l_1 l_2}(|\vec l_1+\vec l_2|) \right. \nonumber \\
          &+ & \left.
          P_{l_2l_4}^{l_1 l_3}(|\vec l_1+\vec l_3|) +
          P_{l_2l_3}^{l_1 l_4}(|\vec l_1+\vec l_4|) \right], \nonumber\\
\label{eqn:localtri}
\end{eqnarray}
where
\begin{eqnarray}
               P_{l_3l_4}^{l_1 l_2}(|\vec l_1+\vec l_2|) &=&
               4 C_{|\vec l_1+\vec l_2|} \left[ C_{l_1} C_{l_3} + C_{l_1}
               C_{l_4} \right. \nonumber \\
               & & \left. + C_{l_2} C_{l_3} + C_{l_2}
               C_{l_4}\right].
\label{eqn:Pdefn}
\end{eqnarray}
Again, the trispectrum is nonvanishing only for $\vec l_1 +\vec
l_2 +\vec l_3+ \vec l_4=0$, that is, only for quadrilaterals in
Fourier space.

\section{Minimum-variance non-Gaussianity Estimators}
\label{sec:estimators}

We now review how to measure $\hnl$ from the
bispectrum and the trispectrum.  To keep our arguments clear
(and since the current goal is simply detection of a departure
from non-Gaussianity, rather than precise evaluation of $\fnl$),
we assume the null hypothesis $\hnl=0$ in the
evaluation of noises and construction of estimators.  The
generalization to nonzero $\hnl$ is straightforward
\cite{Creminelli:2006gc}.

\subsection{The bispectrum}

From Eqs.~(\ref{eqn:bispectrum}) and (\ref{eqn:lmbispectrum}),
each triangle $\vec l_1 +\vec l_2 +\vec l_3 =0$ gives an estimator,
\begin{equation}
     (\hnlb)_{123} = \frac{ T_{\vec l_1} T_{\vec l_2} T_{\vec
     l_3} }{\Omega B(l_1,l_2,l_3)/\hnl},
\label{eqn:onetriangle}
\end{equation}
with variance [using
Eq.~(\ref{eqn:powerspectrum})],\footnote{Here we ignore the
negligible contributions from triangles and for the trispectrum
below, quadrilaterals, where two sides have the same length.  We
do, however, include these configurations in the numerical
analysis described in Appendix \ref{sec:appendixb} and verify that
this assumption is warranted.}
\begin{equation}
     \frac{\Omega^3 C_{l_1} C_{l_2} C_{l_3}}{\left[ \Omega
     B(l_1,l_2,l_3)/\hnl \right]^2}.
\label{eqn:singlevariance}
\end{equation}
The minimum-variance estimator is constructed by adding all of
these estimators with inverse-variance weighting.  It is
\begin{equation}
     \hnlb = \sigma_b^{2} \sum \frac{ T_{\vec l_1} T_{\vec
     l_2} T_{\vec l_3} B(l_1,l_2,l_3)/\hnl}{ \Omega^2
     C_{l_1}C_{l_2}C_{l_3}},
\label{eqn:biestimator}
\end{equation}
and it has inverse variance,
\begin{equation}
     \sigma_b^{-2} = \sum \frac{ \left[ B(l_1,l_2,l_3)/\hnl
     \right]^2}{\Omega C_{l_1}C_{l_2}C_{l_3}}.
\label{eqn:binoise}
\end{equation}
The sums in Eqs.~(\ref{eqn:biestimator}) and (\ref{eqn:binoise})
are taken over all {\it distinct} triangles with $\vec l_1 + \vec l_2 +\vec
l_3=0$.  We may then take $\vec L\equiv \vec l_3$ to be the
shortest side of the triangle---i.e., $l_1,l_2 > L$---and
re-write the estimator as,
\begin{eqnarray}
     \hnlb &=& \frac{1}{2}\sigma_b^{2} \sum_{\vec L} \frac{1}{C_L}
     \nonumber \\
     & \times &
     \sum_{\vec l_1+\vec l_2=-\vec L,\,l_1,l_2>L} \frac{
     T_{\vec l_1} T_{\vec l_2} T_{\vec L} B(l_1,l_2,L)/\hnl}{
     \Omega^2 C_{l_1}C_{l_2}}, \nonumber \\
\label{eqn:biestimatorrewrite}
\end{eqnarray}
and the inverse-variance as
\begin{equation}
     \sigma_b^{-2} =  \frac{1}{2}\sum_{\vec L} \frac{1}{C_L}
     \sum_{
     \vec l_1+\vec l_2 =-\vec L,\, l_1,l_2>L}
     \frac{ \left[ B(l_1,l_2,L)/\hnl
     \right]^2}{\Omega C_{l_1}C_{l_2}}.
\label{eqn:binoise2}
\end{equation}
The factor of $1/2$ is included to account for double counting
of identical triangles, those with $\vec l_1 \leftrightarrow
\vec l_2$.

\subsubsection{Approximation to the Bispectrum Estimator}

Now consider the variance $\sigma_b^2$ with which $\hnl$
can be measured from the bispectrum.  Take $C_l = A/l^2$ for
the power spectrum, where $A\simeq 6\times10^{-10}$ is the
power-spectrum normalization.  The bispectrum in
Eq.~(\ref{eqn:lmbispectrum}) is maximized for squeezed
triangles, those with $L \ll l_1,l_2$, and thus with $l_1 \simeq
l_2$.  In this limit, the bispectrum can be approximated
$B(l_1,l_2,L) \simeq 4 A^2 \hnl L^{-2}l_1^{-2}$.  Then, from
Eq.~(\ref{eqn:binoise2}) the inverse variance (and thus the
signal-to-noise) is dominated by squeezed triangles, and it is
furthermore dominated by those triangles with the modes $\vec L$
of the {\it smallest} magnitudes $L$.

\begin{figure}[htbp]
\centering
\includegraphics[width=0.35\textwidth]{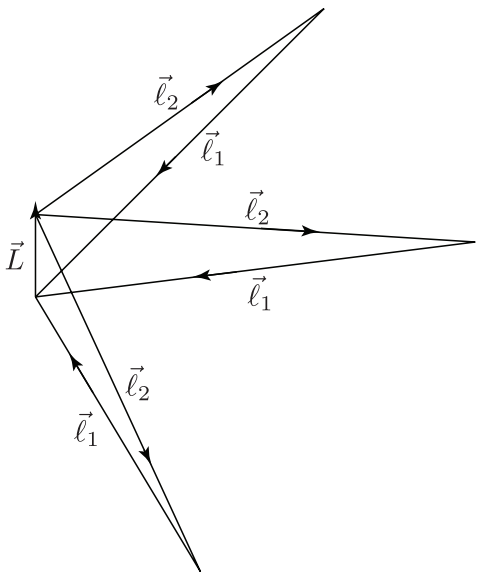}
\caption{Three triangles that all share a shortest side $\vec
     L$.}
\label{fig:triangle}
\end{figure}

More precisely, let us evaluate the contribution
$(\sigma_b^{-2})_{\vec L}$ to the inverse variance obtained from all
triangles that share the same shortest side $\vec L$, as shown
in Fig.~\ref{fig:triangle}.  Since
this contribution is dominated by modes with $\vec l_1\simeq
\vec l_2$, the inverse-variance from these triangles is,
\begin{eqnarray}
     (\sigma_b^{-2})_{\vec L} &\simeq& \frac{1}{2\Omega} \frac{L^2}{A}
     \sum_{\vec l_1} \frac{ (4C_L C_{l_1})^2}{C_{l_1}^2} =\frac{8
     A}{\Omega L^2} \sum_{\vec l} 1 \nonumber \\
     & \simeq& \frac{8 
     A}{L^2} \frac{1}{2\pi} \int_L^{l_{\mathrm{max}}} l \, dl
     \simeq \frac{2 A}{ \pi L^2} l_{\mathrm{max}}^2,
\label{eqn:Dl1estimate}
\end{eqnarray}
where we have used $\sum_{\vec l} = \Omega\int d^2l/(2\pi)^2$ in
the last line.

The full estimator then sums over all $\vec L$ as in
Eq.~(\ref{eqn:biestimatorrewrite}).
The full inverse-variance is then
\begin{eqnarray}
     \sigma_b^{-2} &=& \sum_{\vec L} (\sigma_b^{-2})_{\vec L} =
     \Omega \int\frac{d^2L}{(2\pi)^2} (\sigma_b^{-2})_{\vec L}
     \nonumber \\
     &\simeq& \frac{A \Omega}{\pi^2} l_{\mathrm{max}}^2 \ln
     \frac{L_{\mathrm{max}}}{L_{\mathrm{min}}} \nonumber\\
     &\simeq& \frac{4 A f_{\mathrm{sky}} l_{\mathrm{max}}^2}{\pi} \ln
     \frac{L_{\mathrm{max}}}{L_{\mathrm{min}}},
\label{eqn:fullbivariance}
\end{eqnarray}
in agreement with Ref.~\cite{Babich:2004yc}.

To summarize: (1) the signal-to-noise is
greatly dominated by triangles with one side much shorter than
the other two. (2) The signal-to-noise is dominated primarily by
those with the smallest short side.  (3) The
contribution to the full signal-to-noise is equal per
logarithmic interval of $L$, the magnitude of the smallest mode
in the triangle.  (4)  Even if there is a huge number of
triangles that enter the estimator, the error in the estimator
is still dominated by the cosmic variance associated with the
values of $T_{\vec L}$ for the $\vec L$ modes of the smallest $L$.

Since the variance is dominated by squeezed triangles, we can
approximate the estimator, Eq.~(\ref{eqn:biestimatorrewrite}), as
\begin{equation}
     {\hnle}^b = \frac{2\sigma_b^2}{A\Omega^2 }
     \sum_{\vec L} T_{\vec L} X_{\vec L},
\label{eqn:approxbiestimator}
\end{equation}
where
\begin{equation}
     X_{\vec L} \equiv \sum_{\vec l} T_{\vec l}
     T_{-\vec L - \vec l} l^2.
\label{eqn:Xdefn}
\end{equation}

\subsection{The trispectrum}

Now consider the trispectrum.  Each distinct quadrilateral $\vec
l_1 + \vec l_2+\vec l_3 +\vec l_4=0$ gives an estimator for the
trispectrum with some variance.  Adding the individual
estimators with inverse-variance weighting gives the
minimum-variance estimator,\footnote{Strictly speaking, one must
subtract the connected part of the trispectrum.  We omit this
term to keep our expression compact, but it is included in the
analytic and numerical calculations of the variances and
covariances discussed below.}
\begin{equation}
     \hnlett = \sigma_t^{2} \sum \frac{
     T_{\vec l_1} T_{\vec l_2}
     T_{\vec l_3} T_{\vec l_4} {\cal T}(\vec l_1,\vec l_2,\vec
     l_3,\vec l_4)/\hnl^2}{\Omega^3 C_{l_1}C_{l_2}C_{l_3}C_{l_4}},
\label{eqn:triestimator}
\end{equation}
and the inverse variance,
\begin{equation}
     \sigma_t^{-2} = \sum \frac{ \left[{\cal T}(\vec
     l_1,\vec l_2,\vec l_3,\vec l_4)/\hnl^2 \right]^2}{\Omega^2
     C_{l_1} C_{l_2} C_{l_3} C_{l_4}}.
\label{eqn:trivariance}
\end{equation}
The sums here are over all distinct quadrilateral $\vec l_1
+\vec l_2 +\vec l_3 + \vec l_4=0$, and we again neglect
quadrilaterals where two or more sides are the same.

Each quadrilateral will have a smallest diagonal, which we call
$\vec L$.  The quadrilateral is then described by two triangles
that each share their smallest side $\vec L$; the two sides of
the first triangle will be $\vec l_1$ and $\vec l_2$ and the two
sides of the second triangle will be $\vec l_3$ and $\vec l_4$.
We can then re-write the sums in Eqs.~(\ref{eqn:triestimator})
and (\ref{eqn:trivariance}) as
\begin{equation}
     \sum_{\vec L} \,\, \sum_{\vec l_1+\vec l_2=\vec L} \, \,
     \sum_{\vec l_3+\vec l_4= -\vec L}.
\label{eqn:sums}
\end{equation}
The sum here is only over combinations of $\{\vec l_1,\vec l_2,\vec l_3,\vec l_4\}$ 
where the lengths of the two other diagonals, $|\vec l_1+\vec l_4|=|\vec l_2+\vec l_3|$ 
and $|\vec l_2+\vec l_4|=|\vec l_1+\vec l_3|$, are both $>L$, so that $L$ is the 
shortest diagonal [cf. Eq.~(\ref{eqn:localtri})].

\begin{figure}[htbp]
\centering
\includegraphics[width=0.48\textwidth]{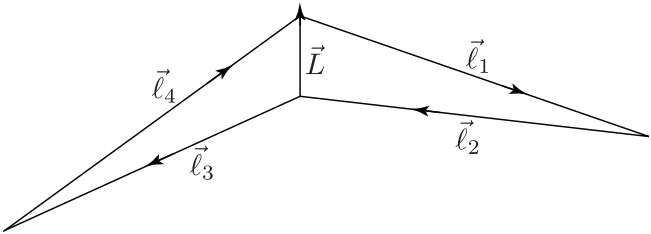}
\caption{An example of an elongated quadrilateral with a
     shortest diagonal $\vec L$.  Note that it is equivalent to
     two elongated triangles that share the same shortest side $\vec
     L$.}
\label{fig:quadrilateral}
\end{figure}

Let's now consider the local-model trispectrum given in
Eqs.~(\ref{eqn:localtri}) and (\ref{eqn:Pdefn}).  The three
terms in Eq.~(\ref{eqn:localtri}) sum over the three diagonals
of the quadrilateral.  Eq.~(\ref{eqn:Pdefn}) then shows that each
of these terms is the product of the power spectrum $C_L$
evaluated for the diagonal (e.g., $\vec L = \vec l_1+\vec l_2 =
-\vec l_3 -\vec l_4$) times a sum of products of power spectra
evaluated for each of the quadrilateral sides.  The
quadrilateral is thus maximized for highly elongated
quadrilaterals, those with $l_i \gg L$, with one short
diagonal, as shown in Fig.~\ref{fig:quadrilateral}.
The trispectrum for these elongated quadrilaterals may be
approximated as ${\cal T}(\vec l_1, \vec l_2,\vec l_3,\vec l_4)
\simeq 16 \hnl^2 C_L C_{l_1} C_{l_3}$.

Now consider the contribution
$(\sigma_t^{-2})_{\vec L}$ to the inverse
variance from all quadrilaterals that share the same shortest
diagonal $\vec L$.  Using Eq.~(\ref{eqn:trivariance}) and
approximating the trispectrum by the squeezed limit, this is
\begin{eqnarray}
     (\sigma_t^{-2})_{\vec L} &\simeq& \frac{1}{8}
     \sum_{\vec l_1} \sum_{\vec l_3}
     \frac{ \left( 16 C_L C_{l_1} C_{l_3} \right)^2}{ \Omega^2 (C_{l_1}
     C_{l_3})^2 } \nonumber \\
     & = & \frac{32\,A^2}{\Omega^2 L^4} \left( \sum_{\vec l} 1
     \right)^2 = \frac{2}{\pi^2} \frac{A^2}{L^4}
     l_{\mathrm{max}}^4.
\label{eqn:Lconttotri}
\end{eqnarray}
The factor $1/8$ in the first line accounts for the $\vec l_1
\leftrightarrow \vec l_2$ and $\vec l_3 \leftrightarrow \vec
l_4$ symmetries and the symmetry under interchange of the $(\vec
l_1,\vec l_2)$ and $(\vec l_3, \vec l_4)$ triangles.
Again, the full variance is obtained by summing over $\vec L$
modes.  Thus, 
\begin{equation}
     \sigma_t^{-2} \simeq \frac{2 f_{\mathrm{sky}}}{\pi^2}
     \frac{A^2}{L_{\mathrm{min}}^2} l_{\mathrm{max}}^4.
\end{equation}
Note that we obtain the $l_{\mathrm{max}}^{-4}$ scaling of
the variance noted in Ref.~\cite{Kogo:2006kh}.  Recall that
$\sigma_t^2$ is a variance to $\hnl^2$ (rather than $\hnl$).
Thus, the ratio of the smallest $\hnl$ detectable via the
trispectrum to the smallest detectable via the bispectrum is
$\sqrt{\sigma_t/\sigma_b^2} \simeq 1.7\, f_{\mathrm{sky}}^{1/4}
\left[ L_{\mathrm{min}} \ln (L_{\mathrm{max}}/L_{\mathrm{min}})
\right]^{1/2}$.  For reasonable
values of $L_{\mathrm{min}}$ and $L_{\mathrm{max}}$, the
smallest $\hnl$ detectable with the bispectrum
is smaller, by a factor of order a few, than that detectable
with the trispectrum
\cite{Hu:2001fa,Okamoto:2002ik,Kogo:2006kh}.

We can now derive an approximation for $\hnlett$
noting that the variance, and thus the signal-to-noise, is
dominated by equilateral triangles.  From
Eq.~(\ref{eqn:triestimator}), and using the squeezed limit for
the trispectrum, we find,
\begin{equation}
     \hnlett = \frac{2}{3} \sigma_t^{2}
     \sum_{\vec L} \frac{1}{L^2} X_{\vec L}^2,
\label{eqn:approxtriestimator}
\end{equation}
where $X_{\vec L}$ is the quantity given in
Eq.~(\ref{eqn:Xdefn}).  Comparing with the estimator,
Eq.~(\ref{eqn:approxbiestimator}), we see that {\it this
estimator is constructed from precisely the same sums of
triangles as the bispectrum estimator}.  Strictly speaking, the
bispectrum estimator for $\hnl$ involves a sum over a huge
number of triangles; the number of such triangles scales as
$N_{\mathrm{pix}}^2/6$ with the number of pixels in the map.
Likewise, the trispectrum estimator for $\hnl^2$ involves a sum
over all quadrilaterals, and the number of these scales as
$N_{\mathrm{pix}}^3/24$.  Thus, one naively expects the
correlation between the estimators to be extremely weak, given
the huge number of bispectrum and trispectrum configurations.
Eqs.~(\ref{eqn:approxbiestimator}) and
(\ref{eqn:approxtriestimator}) show, however, that the quadrilateral configurations that dominate 
the trispectrum estimator for $\hnl^2$ are
very closely related to the triangle configurations that
dominate the bispectrum estimator for $\hnl$.

\section{Correlation between bispectrum and trispectrum
estimators for $\hnl$}
\label{sec:crosscorrelation}

Since the bispectrum and trispectrum estimators for $\hnl$ are
both constructed from the same CMB map, it is expected that
there should be some correlation between the two estimators.
Eqs.~(\ref{eqn:approxbiestimator}) and
(\ref{eqn:approxtriestimator}) help clarify the nature of the
correlation.  Clearly, if we use for the bispectrum
estimator only triangles that share a single shortest side $\vec
L$ and for the trispectrum estimator only quadrilaterals with
the same $\vec L$ as the shortest diagonal, then the two
estimators provide the same quantity, modulo the difference
between the magnitude $|T_{\vec L}|^2$ (from the bispectrum
estimator) and its expectation value $A/L^2$ (from the
trispectrum estimator).  

However, we have not only triangles/quadrilaterals from a
single $\vec L$ shortest side/diagonal, but those constructed
from many $\vec L$'s.  The correlation between the
bispectrum and trispectrum estimators should thus decrease as
the number of $\vec L$ modes increases in the same way that
the means $\VEV{x}$ and $\VEV{x^2}$ measured with a large number
$N$ of data points $x_i$ will become uncorrelated as $N$ becomes large.

Of course since $\hnlb$ is linear in $T_{\vec L}$, the
covariance between $\hnlb$ and $\hnlett$ will be
zero.  However, the correlation between $(\hnlb)^2$ and
$\hnlett$ will be nonzero.  We thus now estimate the
magnitude of the correlation coefficient, which we define as
\begin{equation}
     r\equiv \frac{\VEV{\Delta\left((\hnlb)^2\right)
     \Delta\left(\hnlett\right)}} {
     {\VEV{\left[\Delta\left((\hnlb)^2\right)  \right]^2}^{1/2}
     \VEV{\left[\Delta\left(\hnlett\right) \right]^2}^{1/2}} },
\label{eqn:rdefn}
\end{equation}
where $\Delta(Q) \equiv Q -\VEV{Q}$.  To simplify the equations,
we can drop the prefactors in Eqs.~(\ref{eqn:approxbiestimator})
and (\ref{eqn:approxtriestimator}) and deal with quantities,
\begin{equation}
     F \equiv \sum_{\vec L} T_{\vec L} X_{\vec L}, \qquad
     G \equiv \sum_{\vec L} \frac{A}{L^2} X_{\vec L}^2.
\label{eqn:FGdefn}
\end{equation}
The desired correlation coefficient is then
\begin{equation}
     r = \frac{ \VEV{\Delta(F^2) \Delta G}}{ \VEV{
     \left[\Delta(F^2) \right]^2}^{1/2} \VEV{(\Delta
     G)^2}^{1/2}}.
\label{eqn:rFG}
\end{equation}

We begin by noting that $X_{\vec L}$ is a random variable with
zero mean.  In the large-$l_{\mathrm{max}}$ limit, it will be
well approximated by a Gaussian random variable, in which case
$\VEV{X_{\vec L}^4} =3 \VEV{X_{\vec L}^2}^2$.  Some other useful
relations include,
\begin{equation}
     \VEV{F^2} = \sum_{\vec L_1,\vec L_2} \VEV{T_{\vec L_1}
      T_{\vec L_2} X_{\vec L_1} X_{\vec L_2}} =  \Omega
     \sum_{\vec L} \frac{A}{L^2} \VEV{X_{\vec L}^2},
\end{equation}
\begin{equation}
     \VEV{G} = \sum_{\vec L} \frac{A}{L^2} \VEV{X_{\vec L}^2}
     =\VEV{F^2}/\Omega,
\end{equation}
\begin{equation}
     \VEV{G^2} = \sum_{\vec L_1,\vec L_2}   \frac{A^2}{L_1^2
     L_2^2} \VEV{ X_{\vec L_1}^2 X_{\vec L_2}^2} 
    = \VEV{G}^2 + 2 \sum_{\vec L} \frac{A^2}{L^4} \VEV{
    X_{\vec L}^2}^2,
\end{equation}
\begin{eqnarray}
     \VEV{F^2G} &=& \sum_{\vec L_1}\sum_{\vec L_2}\sum_{\vec
     L_3} \frac{A}{L_1^2} \VEV{ T_{\vec L_2}  T_{\vec
     L_3} } \VEV{X_{\vec L_1}^2 X_{\vec L_2} X_{\vec L_3}}
     \nonumber \\
     &=&\Omega \sum_{\vec L_1,\vec L_2} \frac{A}{L_1^2} \frac{A}{L_2^2}
     \VEV{ X_{\vec L_1}^2 X_{\vec L_2}^2} = \Omega \VEV{G^2}. 
\end{eqnarray}
Also, since $F$ is a sum over (approximately) Gaussian random
variables, it is also well approximated by a Gaussian random
variable, and so $\VEV{F^4} \simeq 3 \VEV{F^2}^2$.

From these relations, it follows that
\begin{equation}
     \VEV{ \Delta(F^2) \Delta G} = \VEV{F^2G}- \VEV{F^2}\VEV{G}
     = \Omega\left[\VEV{G^2}-\VEV{G}^2 \right],
\end{equation}
and thus that
\begin{eqnarray}
     r &=&\frac{\Omega \VEV{(\Delta G)^2}^{1/2}}{\sqrt{2}
     \VEV{F^2}} = \frac { \left( \sum_{\vec L} L^{-4}
     \right)^{1/2}}{ \sum_{\vec L} L^{-2}}
     \nonumber \\
     &=& \left[ 2\sqrt{\pi f_{\mathrm{sky}}} L_{\mathrm{min}}
     \ln\left(L_{\mathrm{max}}/L_{\mathrm{min}} \right)
     \right]^{-1}.
\label{eqn:cccoeff}
\end{eqnarray}
Thus, if $L_{\mathrm{max}}$ is small, then the
correlation will be large.  However, the correlation
coefficient decreases as $[\ln(L_{\mathrm{max}})]^{-1}$, and it
will become negligible in the limit that $L_{\mathrm{max}}$ is
large.

Strictly speaking, the $X_{\vec L}$ are not entirely
statistically independent, as we have assumed here, as many are
constructed from the same measurements.  They are also not
perfectly Gaussian, as we have assumed.  However, as we discuss
in Appendix \ref{sec:appendixb}, we have checked with a full
numerical calculation of the correlation coefficient that the
basic conclusions---and particularly the scaling of the
correlation coefficient $r$ with $L_{\mathrm{max}}$---are sound.

\section{Conclusions}
\label{sec:conclusion}

A large body of recent work has focused on tests of the local
model for non-Gaussianity that can be performed with
measurement of the CMB trispectrum and bispectrum.  Here we have
clarified how the bispectrum and trispectrum may provide
statistically independent information on the local-model
non-Gaussianity parameter $\fnl$, even if the bispectrum
estimator for $\fnl$ saturates the Cramer-Rao bound.  
The basic point is that the Cramer-Rao inequality puts a lower
limit to the variance with which a given parameter can be
measured.  If the likelihood function is precisely Gaussian,
then the likelihood is described entirely by the variance.
However, if the likelihood function is not precisely Gaussian,
then there is more information in the likelihood beyond the
variance (see, e.g., Section VI in Ref.~\cite{Jungman:1995bz}).
In the current problem, this is manifest in that a
statistically-independent measurement of $\fnl^2$ can be
obtained from the trispectrum without contributing to the
variance of $\fnl$.  

We then built on an observation of Ref.~\cite{Creminelli:2006gc} to
illustrate the nature of the correlation between the
bispectrum estimator for $\fnl$ and the trispectrum estimator of
$\fnl^2$.  This analysis demonstrates that the two estimators do
indeed become statistically independent in the
large-$l_{\mathrm{max}}$ limit.

Throughout we have made the null hypothesis $\fnl=0$ to
estimate the variances with which $\fnl$ can be
measured from the bispectrum and with which $\fnl^2$ can be
measured from the trispectrum.  This is suitable if one is
simply searching the data for departures from the null
hypothesis.  However, as emphasized by
Ref.~\cite{Creminelli:2006gc}, the minimum-variance estimators
constructed under the null hypothesis are no longer optimal if
there is a strong signal.  If so, then forecasts of
signal-to-noise made with the null hypothesis are no longer
valid in the limit of large signal-to-noise, and this calls into
question claims \cite{Kogo:2006kh} that the trispectrum
will provide a better probe of the local model in the large-S/N
limit.  In this limit, a new bispectrum
estimator can be constructed to saturate the Cramer-Rao bound
\cite{Creminelli:2006gc}, and an analogous optimal trispectrum
estimator can in principle be found.  Still, the observation
that the bispectrum and trispectrum estimators in the local
model are constructed from the same sums of triangles suggests
that the precisions with which $\fnl$ can be measured, in the
high-S/N limit, from the bispectrum and trispectrum will be
roughly comparable.  

Although we assumed the null hypothesis to argue that the
bispectrum and trispectrum estimators for $\fnl$ are
independent, the same arguments should also apply in the
high-S/N limit.  For example, if the bispectrum estimator finds
$\fnl$ to be different from zero, with best-fit value $\bar
\fnl$, then the likelihood can be re-parametrized in terms of a
quantity $\epsilon=\fnl-\bar\fnl$ that quantifies the departure
from the new null hypothesis $\fnl=\bar\fnl$.  Measurement of $\epsilon$ with
the trispectrum can then be used to provide a statistically
independent consistency check of the model. Or, in simpler
terms, the skewness and kurtosis are still two statistically
independent quantities that can be obtained from a measured
distribution, even if the skewness (or kurtosis) of that
distribution is nonzero.

Throughout, we have made approximations and simplifications to
make the basic conceptual points clear, and we have restricted
our attention simply to the local model, which we have here
defined to be $\Phi=\phi+\fnl(\phi^2-\VEV{\phi^2})$.  However,
inflationary models predict a wider range of trispectra \cite{others}.
Likewise, analysis of real data will introduce a number of
ingredients that we have excised from our simplified analysis.
Still, we hope that the points we have made here may assist in the
interpretation and understanding of experimental results and
perhaps elucidate statistical tests of other, more general,
non-Gaussian models.

\begin{acknowledgments}
MK thanks the support of the Miller Institute for Basic Research
in Science and the hospitality of the Department of Physics at
the University of California, Berkeley where part of this work was
completed.  MK was supported at Caltech by DoE
DE-FG03-92-ER40701, NASA NNX10AD04G, and the Gordon and Betty
Moore Foundation.
\end{acknowledgments}

\appendix
\section{The continuum-discretuum connection}
\label{sec:appendix}

In this paper we have chosen to work with discrete Fourier
transforms where the calculations of variances and covariances
are more straightforward.  Here we show how to derive the
expressions for power spectra, bispectra, and trispectra for
this discrete formalism to the continuum analysis discussed in
most of the theoretical literature.

Following Ref.~\cite{Creminelli:2006gc}, we note that
\begin{equation}
     T_{\vec l} = \int\, d^2\vec \theta\, e^{-i\vec l\cdot
     \vec \theta} T(\vec\theta) \simeq
     \frac{\Omega}{N_{\mathrm{pix}}} \sum_{\vec\theta} e^{-i\vec l\cdot
     \vec \theta} T(\vec\theta),
\label{eqn:Tlcorrespondence}
\end{equation}
where $\Omega = 4\pi f_{\mathrm{sky}}$ is the area of sky (in
steradians) surveyed, from which we infer the correspondence
$\sum_{\vec\theta} \Leftrightarrow (N_{\mathrm{pix}}/\Omega)
\int d^2\vec\theta$.  Likewise,
\begin{equation}
     T(\vec \theta) = \int\, \frac{d^2\vec l}{(2\pi)^2}
     \, e^{i\vec l\cdot \vec \theta}  T_{\vec l} \simeq
     \frac{1}{\Omega} \sum_{\vec l} e^{-i\vec l\cdot
     \vec \theta} T(\vec\theta),
\label{eqn:Tthetacorrespondence}
\end{equation}
from which we infer the correspondence $\sum_{\vec l}
\Leftrightarrow \Omega \int d^2\vec l/(2\pi)^2$.
The Dirac delta function is then written in the discrete
formalism as a Kronecker delta as follows:
\begin{equation}
     (2\pi)^2 \delta(\vec l-\vec l') = \int\, d^2\vec\theta
     e^{i\vec\theta \cdot (\vec l-\vec l')} \simeq
     \frac{\Omega}{N_{\mathrm{pix}}} \sum_{\vec\theta}
     e^{i\vec\theta \cdot (\vec l-\vec l')} =\Omega \delta_{\vec
     l,\vec l'}.
\label{eqn:kronecker}
\end{equation}
The definitions in Section~\ref{sec:definitions} of the power
spectrum, bispectrum, and trispectrum follow from this relation.

One advantage of this formulation is that equations can be
checked for consistency using dimensional analysis.  Recalling
that $\theta$ has units $[\theta^2]=$sterad and that temperature has
units $[T(\vec\theta)]$=K, it follows, for example, that
$[T_{\vec l}]=$K-sterad, $[C_l] =$K$^2$-sterad,
$[\hnl]=$K$^{-1}$, $[B]=$K$^3$-sterad$^2$, and $[{\cal
T}]=$K$^4$-sterad$^3$.  As another check, the variance and
covariances should have an appropriate scaling with
$f_{\mathrm{sky}}$ if factors of $\Omega$ are carried properly
through the calculation.

\section{Full correlation between trispectrum and
bispectrum estimators}
\label{sec:appendixb}

As discussed in the text, the minimum-variance bispectrum and
trispectrum estimators for $\hnl$ are given by
\begin{eqnarray}
     \hnlb &=& \sigma_{\rm b}^2\sum_{\vec l_1 + \vec l_2 + \vec
     l_3 =0} \frac{B(l_1,l_2,l_3)}{3! \Omega^2 C_{l_1} C_{l_2}
     C_{l_3}} T_{\vec l_1} T_{\vec l_2} T_{\vec l_3}, \\
     \hnlett &=& \sigma_{\rm t}^2 \sum_{\vec l_1 + \vec l_2 +
     \vec l_3+\vec l_4 =0} \frac{\mathcal{T}(\vec l_1, \vec l_2,
     \vec l_3 , \vec l_4)}{4!\Omega^3C_{l_1}C_{l_2} C_{l_3} C_{l_4}} \\
& \times& T_{\vec l_1} T_{\vec l_2} T_{\vec l_3}  T_{\vec l_4}\nonumber.
\end{eqnarray}
where $\sigma_{\rm b,t}^2$ are the variances of the bispectrum and
trispectrum estimator.  Here we sum over all triangles and
quadrilaterals (not just those with no equal sides), and the
factors of $3!$ and $4!$ take into account double counting of
degenerate terms in the sum and permutation factors for
triangles and quadrilaterals with equal sides.
In Sec.~\ref{sec:crosscorrelation}
we used the squeezed-limit approximation to estimate the
correlation coefficient between $( \hnlb )^2 $ and $\hnlett$.
In this  Appendix we derive the full expression for this
correlation coefficient and verify that the approximations made
in Sec.~\ref{sec:crosscorrelation} are valid.  

The covariance will consist of a weighted sum of the 10-point
function.  However, because of the fact that no two indices in
the trispectrum or each bispectrum estimator can add to zero we
know that two of the bispectrum indices must combine.  The rest
of the covariance will then be diagonal leading to
\begin{eqnarray}
     &&\VEV{( \hnlb )^2\hnlett} = \frac{\sigma_b^4 \sigma_t^2}{4} \nonumber  \\
     && \times
     \sum_{\vec l_1 + \vec l_2=-\vec L, \vec l_3 + \vec l_4 =
     \vec L} \frac{B(L, l_1,l_2) B({L, l_3, l_4}) 
     \mathcal{T}({\vec l_1, \vec l_2, \vec l_3 , \vec
     l_4})/\hnl^4}{\Omega^2 C_L C_{l_1} C_{l_2} C_{l_3}
     C_{l_4}}.  \nonumber \\
\label{eq:BTcross}
\end{eqnarray}

Finally, we need to compute the variance of $( \hnlb )^2$.  To
do this we must compute the 12-point function
\begin{equation}
     \VEV{T_{\vec l_1} T_{\vec l_2} T_{\vec l_3}|
     T_{\vec l_4} T_{\vec l_5} T_{\vec l_6}| T_{\vec t_1}
     T_{\vec t_2} T_{\vec t_3}| T_{\vec t_4} T_{\vec t_5}
     T_{\vec t_6}},
\end{equation}
where all temperatures within each group of three separated by a
`$|$' have zero covariance.  The variance takes the form
\begin{eqnarray}
     \VEV{{\left(( \hnlb )^2\right)^2}} &=&3 \sigma_b^4\\ 
     &+&\sigma_b^8\sum_{\{\vec l,\vec t\}}
     \frac{\mathcal{B}^{\vec l_1 \vec l_2 \vec l_3}_{\vec t_1
     \vec t_2  \vec t_3}}{\Omega^2 C_{l_1}
     C_{l_2}C_{l_3}C_{t_1}C_{t_2}C_{t_3}}, \nonumber
\label{eq:varbsq}
\end{eqnarray}
where
\begin{eqnarray}
     \mathcal{B}^{\vec l_1 \vec l_2 \vec l_3}_{\vec t_1 \vec t_2
     \vec t_3}&\equiv& B({l_1, l_2, l_3}) B ({l_1, t_1, t_2})
     B({l_2, t_1, t_3}) B({l_3, t_2, t_3}) \nonumber\\ 
     &\times& \delta_{\vec l_1+\vec l_2+\vec l_3,0}\delta_{-\vec
     l_1+\vec t_1+\vec t_2,0} \delta_{-\vec l_2-\vec t_1+\vec
     t_3,0} \delta_{\vec l_3+\vec t_2+\vec t_3,0} \nonumber \\
     &+& \frac{3}{2} B(l_1, l_2,l_3)B({l_1 l_2 t_1}) B({t_2,
     t_3, l_3}) B({t_2 ,t_3, t_1})\nonumber \\
     &\times& \delta_{\vec l_1+\vec l_2+\vec l_3,0}\delta_{-\vec
     l_1-\vec l_2+\vec t_1,0} \delta_{\vec t_2+\vec t_3-\vec
     l_3,0} \delta_{\vec t_1+\vec t_2+\vec t_3,0}.\nonumber\\
\end{eqnarray}
Numerically evaluating the sum in Eq.~(\ref{eq:varbsq}) shows
that for $l_{\rm max} \gtrsim 100$ the second (non-Gaussian)
term contributes less than 1\% to the variance of $( \hnlb )^2$.
We have moreover numerically evaluated the exact expression for
the correlation coefficient and verified that, as our estimates
indicate, the correlation is of order $\lesssim 10\%$ for
$l_{\rm min} = 2$ and $l_{\rm max} \gtrsim 100$.

\end{document}